\documentclass[aps,prl,twocolumn,amsmath,amssymb]{revtex4-1}

\usepackage{graphicx}
\usepackage{dcolumn}
\usepackage{bm}
\usepackage{url}

\begin{document}

\title{Point-Defect Optical Transitions and Thermal Ionization Energies \\
from Quantum Monte Carlo Methods: Application to F-center Defect in MgO}

\author{Elif Ertekin}
\altaffiliation{Currently at University of Illinois, ertekin@illinois.edu}
\author{Lucas K. Wagner}
\altaffiliation{Currently at University of Illinois, lkwagner@illinois.edu}
\author{Jeffrey C. Grossman}
\email[e-mail:]{jcg@mit.edu}
\affiliation{Dept. of Materials Science \& Engineering, Massachusetts Institute of Technology, Cambridge, MA 02139}

\date{\today}

\begin{abstract}
\noindent We present an approach to calculation of point defect optical and thermal ionization energies 
based on the highly accurate quantum Monte Carlo methods.  The use of an inherently many--body theory that directly treats electron correlation offers many improvements over the typically-employed density functional theory Kohn-Sham description.  In particular, the use of quantum Monte Carlo methods can help overcome the band gap problem and obviate the need for ad-hoc corrections.  We demonstrate our approach to the calculation of the optical and thermal ionization energies of the F-center defect in magnesium oxide, and obtain excellent agreement with experimental and/or other high-accuracy computational results.  
\end{abstract}

\pacs{}

\maketitle

From electronics to optoelectronics to photovoltaics, point defects influence and even dominate the properties of semiconducting materials~\cite{Chadi:1988vd,Zhang:1990td,VandeWalle:1989uc,Zhang:1998tk,VandeWalle:2004bk,Park:2002tp}.  Quantitative descriptions of the effect of point defects on electronic, optical, and transport properties is critical to  enabling point-defect engineering for  materials design.  However, accurate prediction of point-defect energetics, thermal ionization energies, and optical transition energies from first principles remains a challenge.  Currently, the most widely-used approach based on conventional density functional theory (DFT) suffers from poor descriptions of band gaps that render difficult the accurate description of mid-gap defect states~\cite{Lany:2008gk,Rinke:2009db,Drabold:2007vc,VandeWalle:2004bk}.  Here we demonstrate that, by contrast, an inherently many-body approach based on quantum Monte Carlo (QMC) methods~\cite{Petruzielo:2012bl,Foulkes:2001td} can eliminate these problems and enable high-accuracy calculations of point defect optical and thermal ionization energies.  Our computed optical transition energies are in excellent agreement with experimental and/or other high-accuracy computational results for the same system~\cite{Rinke:2012ip}, and demonstrate that QMC can obtain quantitatively accurate descriptions.  

QMC methods comprise a suite of stochastic tools that enable calculations of material properties based on the many-particle Schr\"{o}dinger equation.  Because of their direct treatment of electron correlation, QMC methods are among the most accurate electronic structure approaches available today, and demonstrate a long and distinguished record of ground-breaking and benchmarking calculations~\cite{Petruzielo:2012bl,Foulkes:2001td,Grossman01}.  In comparison to the other ``beyond-DFT" techniques that are currently explored for calculation of point defect properties (DFT+U, hybrid DFT, and the GW method), QMC is directly based on the true many-body Schr\"{o}dinger equation and offers the possibility of parameter-free accurate band gaps and total energies.  The application of QMC techniques to point defects in solids is still a relatively new field.  To date, a handful of studies have been carried out to compute defect formation energies: interstitials in silicon~\cite{Batista:2006kn,Hood:2003cy,Parker:2010dq}, vacancies in diamond~\cite{Hood:2003cy}, the Schottky defect in MgO~\cite{Alfe:2005eu}, and vacancies and interstitials in aluminum~\cite{Hood:2012jg}.  

\begin{figure}[b]
\includegraphics[width=8.2 cm]{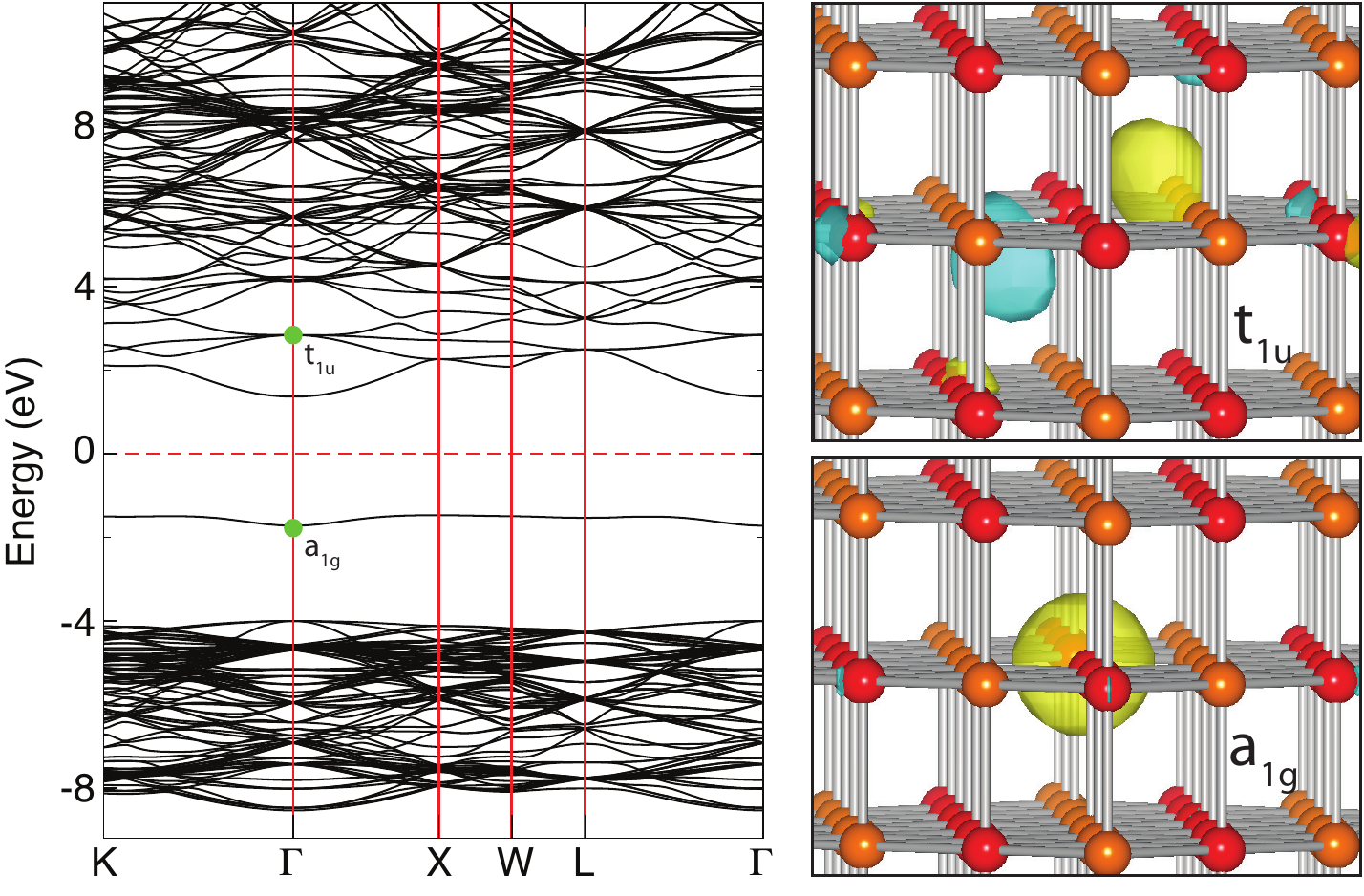}
\caption{\label{fcenter} (Color online).  Left: The electronic band structure of a 64 atom MgO supercell containing a single oxygen vacancy, calculated within DFT-PBE.  The neutral oxygen vacancy introduces a localized mid gap defect level of symmetry a$_{1g}$.  There is also a triply degenerate excited defect level in the conduction band of t$_{1u}$ symmetry.  Right:  The corresponding a$_{1g}$ and t$_{1u}$ Kohn-Sham states plotted at the $\Gamma$-point, showing the localized nature in the vicinity of the vacancy.}
\end{figure}

\begin{table}\footnotesize
\begin{tabular}{ l | c | c | c}
  & DFT-PBE & DMC & Exp~[\onlinecite{NISTwebbook,Whited:1969tq,CRC}] \\ \hline \hline
Lattice const ({\AA}) & 4.25 & 4.22 & 4.216  \\ \hline
Coh. En. (eV/MgO) & 9.50 & 10.18(5) & 10.5  \\ \hline
OP (eV) & 4.83 & 7.96(6) & 7.78 \\ \hline
QP (eV) & -- & 7.9(1) & 7.84 \\ \hline
IP (eV) & -- & 3.28(7) & -- \\ \hline
EA (eV) & -- & 11.17(7) & -- \\ \hline
\end{tabular}
\caption{\label{solid} Comparison of lattice constant, atomization energy, and band gap in MgO solid according to DFT, DMC, and in experiment.  All computed results are obtained using Ne-core pseudo potentials for magnesium.  The optical band gap is determined in DFT from the Kohn-Sham levels, and in DMC from the extrapolated optical excitation energy.  Error bars are shown in parenthesis. Note that the IP and EA are given with respect to the average potential in the supercell.}
\end{table}
 
In this letter, we illustrate the application of the QMC method to the F-center defect in magnesium oxide (MgO) by computing defect formation energies, thermal ionization levels, and optical ionization energies (although the approach is general and can be extended to other materials of interest).  The F-center defect (oxygen vacancy) in MgO is a typical example of an intrinsic point defect in a binary ionic compound~\cite{Jeffries:1982ur,Rosenblatt:1989wn,Chen:1969uo,Kappers:1970tj,KEMP:1969gi,Chen:1990im}.  Despite its apparent simplicity, its properties as deduced from optical absorption and luminescence studies have proven somewhat ambiguous.  Experimental characterization of the F-center in its neutral (F$^0$) and singly ionized (F$^{+1}$) state has been complicated by their nearly identical optical absorption energies~\cite{KEMP:1969gi,Kappers:1970tj,Chen:1990im}.  These energies have been corroborated by recent GW calculations~\cite{Rinke:2012ip}; however, these calculations also predicted optical emission energies that are substantially different from the assigned experimental values, causing the authors to suggest a reinvestigation of the experimental observations.  A particularly compelling possibility is to explore the F-center defect using distinct high-accuracy first-principles techniques to compare results.  Our results - calculated independently using QMC methods - corroborate the GW results and further invite reassessment of the experimental data for the optical emission.  

We first compute the properties of the F-center defect in MgO within a DFT~\cite{Hohenberg:1964ut,Kohn:1965ui} framework as implemented in the SIESTA code~\cite{Artacho:2008jv}, employing the Perdew-Burke-Ernzerhof~\cite{Perdew:1996ug} approximation to the exchange correlation potential.  The inner core electrons are represented by Troullier-Martins pseudopotentials (leaving the Mg $3s$ and O $2s$, $2p$ electrons in valence), and the Kohn-Sham orbitals are represented by a linear combination of numerical pseudo atomic orbitals expanded in a triple-$\zeta$ with polarization Gaussian basis set.  For bulk rocksalt MgO, in agreement with previous DFT calculations~\cite{Marinelli:2003vl,Schleife:2006ez} we find a lattice parameter of 4.25 $\AA$ (4.22 $\AA$ in experiment~\cite{NISTwebbook}), an atomization energy of 9.50 eV/MgO (10.50 eV/MgO in experiment~\cite{NISTwebbook}), and a direct band gap of 4.83 eV at the $\Gamma$-point (a considerable underestimate of the experimental band gap of 7.78 eV~\cite{Whited:1969tq}).  

The DFT band gap underestimate has a severe consequence on the prediction of mid-gap defect states, defect energetics (particularly for occupied defect levels), and defect-induced optical absorption and emission energies.  Broadly, the band gap underestimate arises because in typically-employed mappings of the interacting many-body Schr\"{o}dinger equation to the DFT single-particle ``effective"-potential Kohn-Sham equations, each electron also interacts with itself (self-interaction error~\cite{Ruzsinszky:2007ho,Ruzsinszky:2006gq,MoriSanchez:2006gl,Perdew:1982wa,Zhang:1998im}).  This results in an extraneous Coulomb repulsion that overly delocalizes electronic states.  The self-interaction error, in addition to the absence of a derivative discontinuity in the exchange-correlation potential~\cite{Ruzsinszky:2007ho,Ruzsinszky:2006gq,MoriSanchez:2006gl,Perdew:1982wa,Zhang:1998im}, results in underestimated band gaps that have plagued DFT calculations.  

In Fig.~\ref{fcenter}, we show the DFT-computed electronic band structure of a 63 atom MgO supercell containing an F$^0$-center defect (neutral oxygen vacancy).  In agreement with previous DFT calculations~\cite{Rinke:2012ip}, the F$^0$-center introduces a fully-occupied mid-gap defect level of $a_{1g}$ symmetry into the electronic band structure; higher in the conduction band we also find a triply-degenerate excited defect level of $t_{1u}$ symmetry.  

The QMC calculations reported here are computed within fixed node diffusion Monte Carlo (DMC) as implemented in the QWalk code~\cite{Wagner:2009dy}, with single-determinant Slater-Jastrow trial wave functions constructed from the DFT orbitals, variance-minimized Jastrow coefficients, and a time step of 0.01 au.  To establish that our choice of pseudopotentials is reasonable, we first calculated the bond length, electron affinity, and binding energy of the MgO molecule within DMC.  We tested both Ne and He -core pseudopotentials for the Mg atom in the molecule, and found that (although both give good results) the small core pseudopotential gives a somewhat better description (see supplementary information).  This suggests that including the Mg $2s$ and $2p$ electrons improves the description slightly; however, for the solid, the computational cost of the He-core pseudopotential was prohibitive.

To test the properties of the non-defective MgO solid in QMC, we calculate the atomization energy, the optical band gap, and the quasiparticle band gap.  
All energies are calculated using the  extrapolation framework described in Refs~[\onlinecite{Kolorenc:2011hv,Kolorenc:2008gn,Kolorenc:2010fe}]  with supercells containing 16, 32, and 64 atoms.  Further details are provided in the supplementary information.  We calculate the following states: the ground state $E_g$,  the $\Gamma$-point optically excited state $E_{\Gamma\rightarrow\Gamma}$, the positively charged state $E_+$, and the negatively charged state $E_-$.  
The ionization potential (IP), electron affinity (EA), quasiparticle gap (QP), and optical gap (OP)  are given by 
\begin{eqnarray}
IP & = & E_g-E_+ \\
EA & = & E_--E_g \\
QP & = & EA-IP  \\
OP & = & E_{\Gamma\rightarrow\Gamma}-E_g 
\end{eqnarray}
The results are summarized in Table~\ref{solid}, and show excellent agreement overall with the experimental values.  
The slight underestimation of the atomization energy is likely due to the Ne-core pseudopotential, since the MgO molecule showed a similar effect (see supplementary information), while the gap calculations are close to experiment, overestimating the gap slightly.

\begin{table}\footnotesize
\begin{tabular}{  c | c | c | c  } 
					& Mg-Mg ({\AA}) 	& O-O ({\AA})	& Relaxation (eV) \\
\hline \hline perfect	& 2.98 				& 5.96		& - \\
\hline F		 		& 2.99 				& 5.96  		& 0.003 \\
\hline F$^+$	 		& 3.09 				& 5.90 		& 0.545 \\ 
\hline F$^{+2}$	 		& 3.17 				& 5.84 		& 1.182 
\end{tabular}
\caption{\label{configs} DFT-computed lattice relaxations for the F, F$^+$, and F$^{+2}$ center.  The Mg-Mg distance denotes the separation between Mg atoms that neighbor the missing O atom; similarly the O-O distance denotes the separation between O atoms that neighbor the missing O atom.}
\end{table}

\begin{figure}
\includegraphics[width=7 cm]{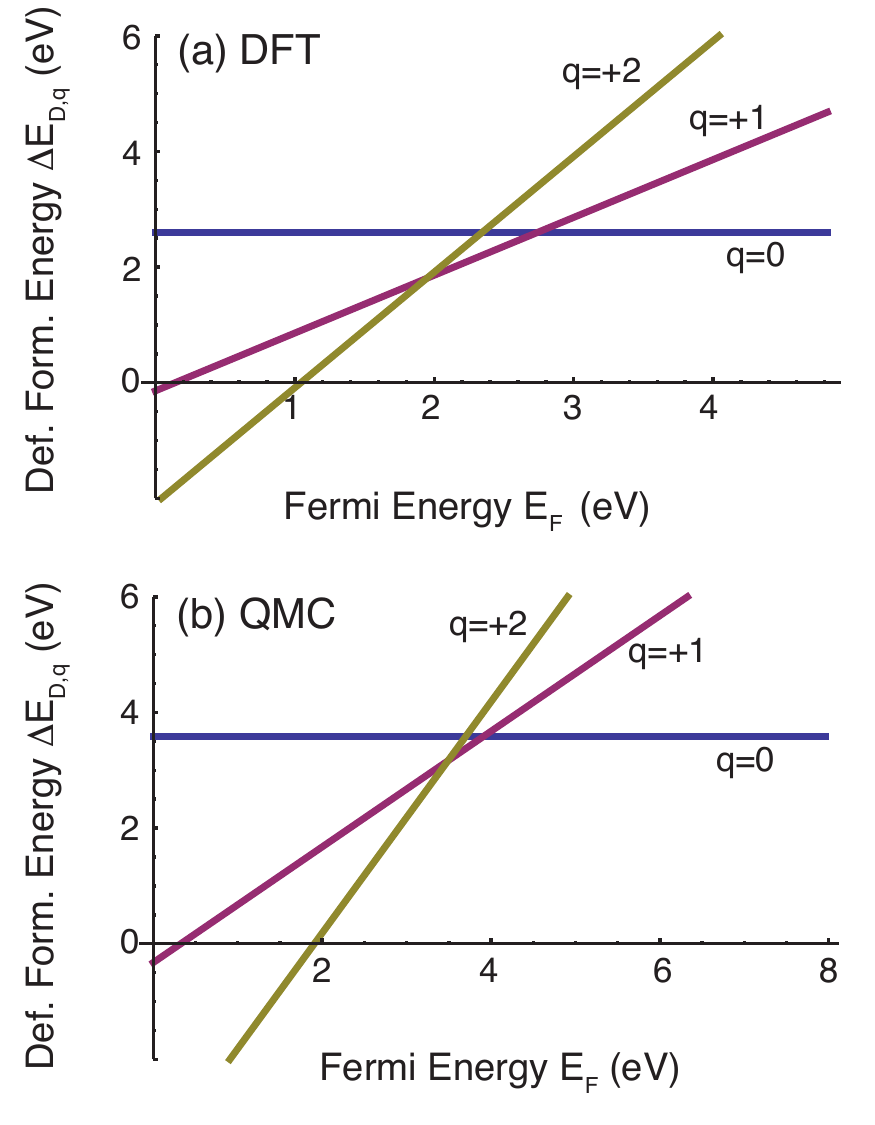}
\caption{\label{thermal} (Color online).  Comparison of F-center defect formation energies and thermal ionization energies in MgO computed in DFT and DMC.  The domain of the Fermi energy (x-axis) is determined by the band gap of the system according to the computational framework; clearly the DMC description of the gap is better and obviates the need for ``band-gap corrections".  In comparison to DFT, DMC modifies somewhat the absolute value of the defect formation energies, but maintains thermal ionization levels near mid gap.  (Note that the error bars of the DMC-computed formation energies are smaller than the line widths in (b).)}
\end{figure}

Now we turn to the F-center defect.  We again use 64 (perfect) and 63 (with F-center) atom supercells, and compute the defect formation energy $\Delta E_{D,q} $ according to
\begin{equation}
\label{defect}
\Delta E_{D,q} = (E_{D,q}-E_{perf})-\sum_i n_i \mu_i + q (E_V + E_F) 
\end{equation}
where $E_{D,q}$ is the (computed) total energy of the supercell containing a defect D in the charge state q, $E_{perf}$ is the (computed) total energy of the perfect supercell, and $n_i$ is the number of atoms of species $i$ added to ($n_i>0$) or removed from ($n_i<0$) the supercell to create the defect~\cite{Zhang:1991ud}.  Different environmental conditions are accommodated by the set of chemical potentials $\mu_i$ for each element by assuming that each is in equilibrium with a physical reservoir such as a gas or a bulk phase.  $E_V$ is the energy of the valence band maximum (the ionization potential in DMC), and $E_F$ is the Fermi energy referenced to $E_V$ so that $0 \le E_F \le E_g$ where $E_g$ is the band gap.  

The thermal ionization energies, which determine the shallow or deep nature of a defect, correspond to the Fermi energies at which the energetically most favored charge state of the defect changes.  According to our DFT calculations, the creation of an F$^0$-center results in the formation of a filled mid-gap defect level (shown in Fig.~\ref{fcenter}).  There is very little lattice relaxation that takes place upon removal of the O, as indicated in Table~\ref{configs}.  However, when an electron is removed from the supercell to form the F$^{+1}$-center, in DFT we find a large lattice relaxation as the positively charged Mg ions move outwards away from, and the negatively charged O ions move inwards towards, the positively charged vacancy in conjunction with a 0.55 eV drop in energy.  Further ionizing the defect into the F$^{+2}$ state in DFT results in further lattice relaxations accompanied by an energy recovery of 1.18 eV.  The DFT defect formation energies obtained from Eq.~\ref{defect} are plotted in Fig.~\ref{thermal}a, showing thermal ionization levels near the middle of the gap.  These formation energies are computed in the Mg-rich limit so that $\mu_{Mg}$ is given by the chemical potential of solid elemental magnesium (and $\mu_{Mg}+\mu_O=\mu_{MgO}$).  In Fig.~\ref{thermal}a, we have used the as-computed DFT band gap for MgO, without correction schemes to artificially open the gap.  Note that we have also ignored the interaction energy between charged defects arising from the use of periodic boundary conditions, which would shift upwards the lines for the formation energy of the F$^{+1}$ and F$^{+2}$ center (and reduce the defect transition levels).  We will ignore this interaction for the QMC calculations as well, and compare both methods on an even footing~\footnote{In principle, the charge defect interaction could be computed using larger supercells and estimating the charged defect interactions by extrapolation, but from a computational standpoint this is prohibitively expensive for QMC.}.  

\begin{figure}[h]
\includegraphics[width=8.6 cm]{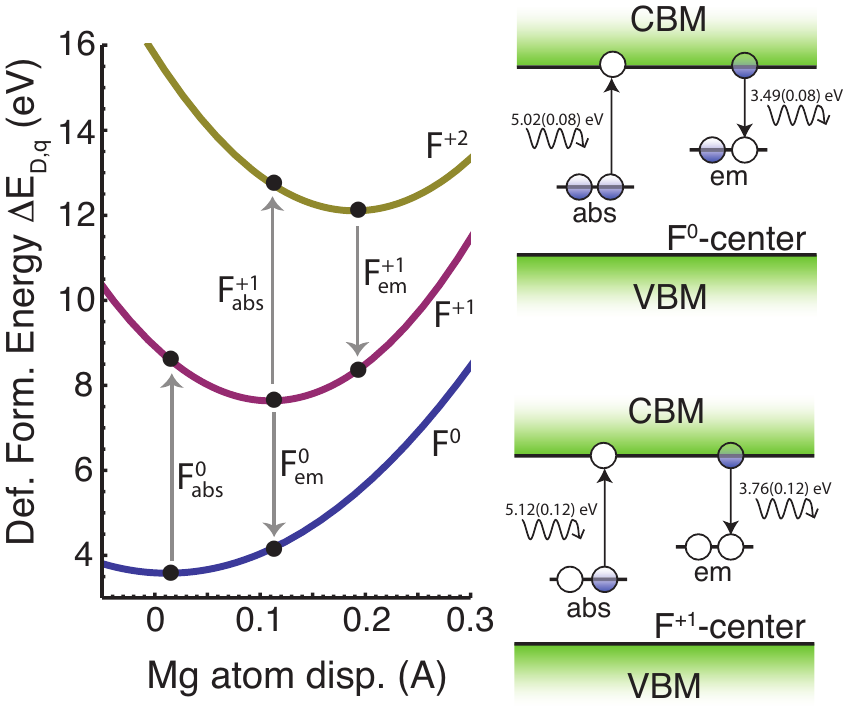}
\begin{tabular}{  l | c | c  | c | c } 
          &  $F_{abs}^0$ & $F_{em}^0$ & $F_{abs}^{+1}$ & $F_{em}^{+1}$ \\
 \hline
 \hline
QMC & 5.0(1) & 3.8(1) & 5.1(1) & 3.5(1) \\ \hline
GW\cite{Rinke:2009db}    & 4.95   & 3.6 & 4.92 & 3.4  \\ \hline
Exp    & 5.00   & 3.1-3.2  & 4.95 & 2.3-2.4 \\ \hline
\end{tabular}

\caption{\label{optical} (Color online).  Optical absorption and emission energies (in eV) computed in DMC for the F-center defect in MgO.  The DMC absorption energies are in excellent agreement with experiment and recently published GW results.  The DMC emission energies are in disagreement with the experimentally assigned values, but match closely to the GW results.}
\end{figure}

For the QMC calculations of total energies of perfect and defected supercells (E$_{perf}$, E$_{D,q}$) and the corresponding thermal ionization levels, we use the relaxed lattice geometries and orbitals obtained from the DFT calculations.  For computational efficiency, only the $\Gamma$ point Kohn-Sham orbitals are used, rather than twist-averaging, since in Eq.~\ref{defect}, the errors in the {\it difference} ($E_{D,q}-E_{perf}$) largely cancel.  For instance, in DFT these errors are 0.26~eV for the F$^0$-center, and $<0.1$~eV for the F$^{+1}$ and  F$^{+2}$ -center, and do not substantively alter our results.  Figure~\ref{thermal}b shows the defect formation energies and the thermal ionization levels as computed within QMC.  The largest single difference between the DFT and the QMC results is, of course, the domain for the Fermi energy $0<E_F<E_g$.  For this material, both DFT and QMC place the transition levels near mid-gap~\footnote{This qualitative similarity is not expected to hold for all material systems, especially for wide gap semiconductors or correlated systems for which the DFT gap error is more insidious.}.  

From Fig.~\ref{thermal}, the overall formation energy of the neutral defect (F$^0$) is higher in QMC by approximately 1.5 eV.  This increase in formation energy may be largely attributed to the fact that DFT-PBE renders the MgO system more delocalized while QMC captures the true ionic nature of the solid.  The overly-delocalized (more metal-like) description in DFT makes the penalty for bond breaking too small, and consequently the defect formation energy too small as well.  This over-delocalization results in the underestimated DFT band gap; the effect on computed energetics are significant for defects with occupied mid-gap defect levels such as the F$^0$ center.  As a result of the underestimated band gap, in DFT the Kohn-Sham level of the deep, doubly occupied F$^0$ level is squeezed too close to the valence band maximum, directly resulting in a low calculated formation energy.  Our finding here is similar to the findings for the formation energies of neutral, interstitial Si atoms in silicon~\cite{Batista:2006kn,Hood:2003cy,Parker:2010dq}, for which DMC calculations show that DFT underestimates formation energies in the case of occupied mid-gap defect levels.  This analysis is also consistent with recent QMC results for defect formation energies in aluminum~\cite{Hood:2012jg} - for metallic systems, for which DFT delocalization problems are less significant, the DMC results are more closely matched.  

For the charged defects F$^{+1}$ and F$^{+2}$, the comparison between the DFT and QMC results is more complex.  The difference between the QMC and DFT -computed formation energies arises from an interplay between the difference in predicted band gap, the occupation of the Kohn-Sham defect level (zero, one, and two electrons for the F$^{+2}$, F$^{+1}$, and F$^0$ center respectively), and the fact that the charged-defect interactions arising from the use of periodic boundary conditions are likely different for the two schemes.  That is, since the degree of localization/delocalization is different, screening is also expected to be different for the two methodologies - most likely enhanced in the case of DFT.  

We now turn to the QMC description of the optical ionization energies, corresponding to vertical Franck-Condon transitions on a configuration coordinate diagram as illustrated in Fig.~\ref{optical}.  An optical transition occurs when a photon is absorbed or emitted by the defect; because this transition essentially takes place instantaneously on the scale of lattice relaxations it occurs at fixed atomic coordinates (and hence is represented as a vertical transition).  Such a transition places the system in an excited vibrational state; for example, F$^0$-center absorption illustrated in Fig.~\ref{optical} refers to the absorption of a photon and the promotion of an electron from the filled mid-gap level to the conduction band, leaving behind an electron in the conduction band and an F$^{+1}$ center in an excited vibrational state (which soon decays to the F$^{+1}$ vibrational ground state).  Therefore, we compute the optical transitions by using the relaxed coordinates of the initial state, and occupying the Kohn-Sham orbitals as appropriate to describe the vibronic state.

Our DMC absorption energies (Fig~\ref{optical})  are in excellent agreement with experiment and remarkably close to the GW-computed values, demonstrating the high-accuracy potential of the DMC methodology.  The DMC emission energies are also remarkably similar to the GW-computed values, but in disagreement with the experimental numbers.  The disagreement between the GW and experimental values led the authors in \cite{Rinke:2009db} to suggest that the low energy signal around 2.3-2.4 eV that is observed in fact arises when electrons in the defect level recombine with holes in the valence band.  We find it notable that two distinct many-body approaches (namely QMC and GW) have yielded similar results for the optical emission transitions in question.  

This leads us to suggest two possibilities.  First, we find it likely that, as suggested by the authors in Ref.~\cite{Rinke:2009db}, the original emission-peak assignment should be revisited.  A second possibility is based on the fact that both our QMC and the GW results are built from DFT-relaxed atomic geometries (Table~\ref{configs}).  It is possible that the GW and QMC results compare favorably because both methods are using similar DFT-relaxed lattice geometries.  If the relaxations are not properly described in DFT, then the many-body energies may be similar but incorrect.  However, the possibility that the lattice geometries are problematic seems unlikely given the exceptional agreement with experiment for the absorption transitions.  

In conclusion, we demonstrate the application of quantum Monte Carlo methods to the calculation of the thermal and optical ionization energies of point defects in solids.  The striking agreement between two highly accurate methods, quantum Monte Carlo and GW, suggests that predictive calculations of point defect properties are now in reach.  Due to its inherently many-body approach and accurate treatment of electron correlation, quantum Monte Carlo shows large promise for the quantitative first-principles calculation of point defect properties.  

We gratefully acknowledge fruitful discussions with S.B. Zhang, Y.Y. Sun, P. Zhang, and T. Abtew.  This work was supported by DOE grant DE-SC0002623.  Calculations were performed in part at the National Energy Research Scientific Computing Center of the Lawrence Berkeley National Laboratory and in part by the National Science Foundation through TeraGrid resources provided by NCSA under grant TG-DMR090027.

\clearpage

\noindent {\bf Supplementary Information} \\

\noindent {\bf I.  Properties of MgO Molecule as Described in Diffusion Monte Carlo} \\

We compute the properties of the MgO molecule in diffusion Monte Carlo (DMC) using both small (helium) and large (neon) -core pseudopsotentials; the results are given in Table~\ref{molecule}.  While both pseudopotentials give excellent agreement to the experimental bond length and electron affinity, we find some difference for the binding energy and the ionization energy.  Using a He-core, rather than Ne-core, pseudopotential  for Mg increases the molecular binding energy from 2.28$\pm$0.01 eV to 2.43$\pm$0.01 eV, in comparison to the experimental value of 2.54$\pm$0.22 eV~\cite{Operti:1989vq,Kim:2001it,NISTwebbook,Recio:1993un}.  This suggests that allowing the Mg $2s$ and $2p$ electrons to participate in the bonding allows more recovery of the binding energy.  We also find that using the Ne-core pseudopotential introduces a small 0.04(2) eV error in the ionization energy compared to the He core pseudopotential.  

\begin{table}[h]\footnotesize
\begin{tabular}{  l | c | c  | c  } 
						& bond   			& binding  		& electron \\
						& length ({\AA})		& energy (eV)		& affinity (eV) \\
\hline \hline DMC, Ne-core PP 	& 1.75 			& 2.28(1)		& 1.76(1)) \\
\hline DMC, Ar-core PP 	& 1.75 			& 2.43(1)		& 1.72(1)\\
\hline Exp. (Refs.~[\onlinecite{Operti:1989vq,Kim:2001it,NISTwebbook,Recio:1993un}])	& 1.75 		& 2.56(21) & 1.630(25)  \\ 
\end{tabular}
\caption{\label{molecule} Comparison of bond length, binding energy, and electron affinity for the MgO molecule according to DMC and in experiment.  Two sets of DMC results are provided, corresponding to the use of neon (large) and helium (small) -core pseudopotentials.  Error bars are shown in parenthesis. }
\end{table}

\noindent {\bf II.  Properties of Crystalline MgO as Described in Diffusion Monte Carlo} \\

\begin{figure}[h]
\includegraphics[width=7.3 cm]{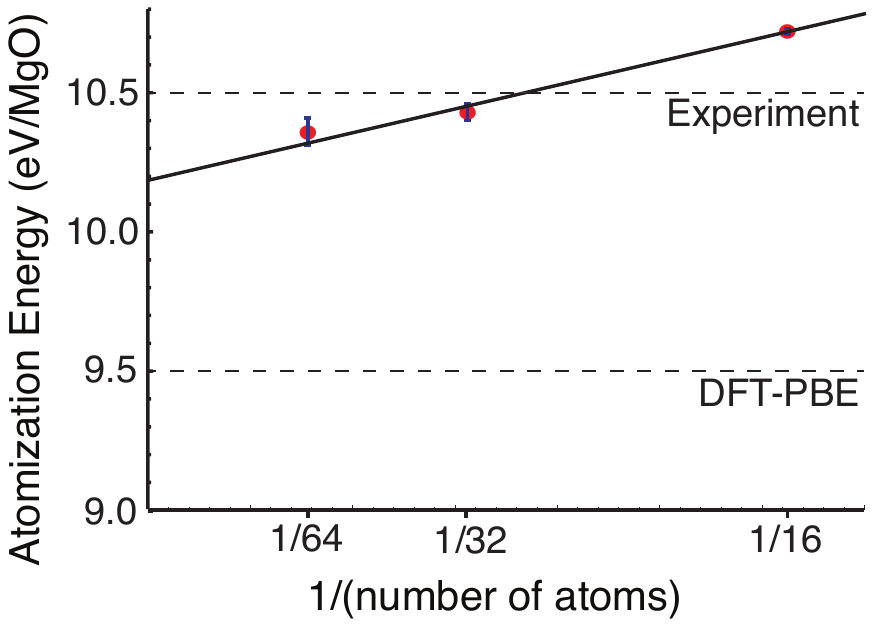}
\caption{\label{atomization} (Color online).  The extrapolated value of the atomization energy as computed in DMC is 10.18$\pm$0.05 eV, in comparison to the experimental and DFT values of 10.5 eV and 9.48 eV, respectively~\cite{NISTwebbook,Whited:1969tq,CRC}.}
\end{figure}

The extrapolation framework described in Refs.~[\onlinecite{Kolorenc:2011hv,Kolorenc:2008gn,Kolorenc:2010fe}] is used to compute the atomization energy for the MgO solid, as shown in Fig.~\ref{atomization}.  We use supercells containing 16, 32, and 64 atoms using twist-averaged boundary conditions.  The dependence of the binding energy on the supercell size reflects the spurious electron correlation that appears in many-body theories when periodic boundary conditions are applied.  This spurious correlation disappears in the infinite size supercell limit.  The results presented in Fig.~\ref{atomization} are computed using twist--averaged boundary conditions.  The extrapolated value of the atomization energy in DMC is 10.18$\pm$0.05 eV per formula unit, in comparison to the experimental and DFT values of 10.5 eV and 9.48 eV, respectively~\cite{NISTwebbook,Whited:1969tq,CRC}.  Although DMC improves the atomization energy in comparison to DFT, it is most likely necessary to include the Mg $2s$ and $2p$ electrons in valence to obtain atomization energies closer to experiment.  

\begin{figure}[h]
\includegraphics[width=7.3 cm]{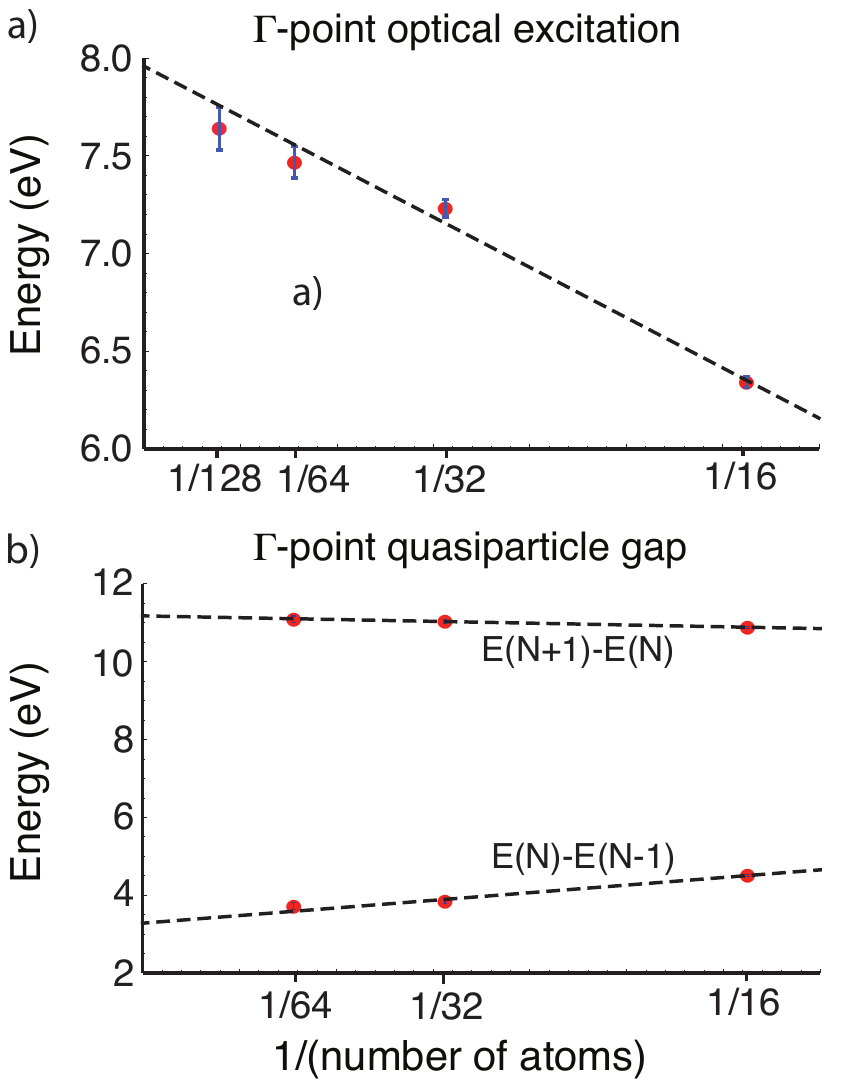}
\caption{\label{gaps} (Color online).  Optical and quasiparticle gap in MgO obtained by extrapolating supercells to infinite size.  The data points for finite-size supercells are computed with DMC.  The extrapolated value of the optical gap is 7.96$\pm$0.06 eV.  The extrapolated ionization potential IP=E(N)-E(N-1)=3.28$\pm$0.07 eV.  The extrapolated electron affinity EA=E(N+1)-E(N)=11.17$\pm$0.07 eV.  This gives a quasiparticle gap of QP=EA-IP=7.89$\pm$0.10~eV.  (Note the EA and IP are referenced to the average electrostatic potential in the supercell.)}
\end{figure}

The optical gap, ionization potential, and electron affinity are similarly computed by extrapolation, as illustrated in Fig.~\ref{gaps}.  Here we use only the $\Gamma$-point orbitals to construct the many-particle wave function, to save computational cost and since the MgO solid exhibits a direct band gap at the $\Gamma$-point.  For these quantities, additional finite size effects are present including (1) periodic image interactions between the electron--hole pair for the optical gap and (2) the electrostatic interaction between charged supercells in the calculation of the ionization potential and electron affinity.   Extrapolating to the infinite supercell limit, we obtain an optical gap of 7.96$\pm$0.06 eV and a quasiparticle gap of 7.89$\pm$0.10~eV, in close agreement with the experimental band gap of 7.8 eV.  

\clearpage

\end{document}